\documentclass[aps,twocolumn,nofootinbib,longbibliography]{revtex4-1}
\usepackage[babel]{csquotes}
\usepackage{graphicx}
\usepackage{amsmath,amssymb}
\usepackage[colorlinks,citecolor=blue,linkcolor=blue,urlcolor=blue]{hyperref}
\usepackage{mathrsfs}
\usepackage{enumerate}
\def\be{\begin{equation}}
\def\ee{\end{equation}}
\def\H{{\cal H}}
\def\tr{{\rm tr}}

\begin{document}

\title{Is Time's Arrow Perspectival?}

\author{Carlo Rovelli}

\affiliation{Aix Marseille Universit\'e, CNRS, CPT, UMR 7332, 13288 Marseille, France\\
Universit\'e de Toulon, CNRS, CPT, UMR 7332, 83957 La Garde, France.}

\date{\small\today}

\begin{abstract}
\noindent 
We observe entropy decrease towards the past.  Does this imply that  in the past the world was in a  non-generic microstate? I point out an alternative. The subsystem to which we belong interacts with the universe via a relatively small number of quantities, which define a coarse graining. Entropy happens to depends on coarse-graining. Therefore the entropy we ascribe to the universe depends on the peculiar coupling between us and the rest of the universe. Low past entropy may be due to the fact that \emph{this coupling} (rather than microstate of the universe) is non-generic. I argue that for \emph{any} generic microstate of a sufficiently rich system there are \emph{always} special subsystems defining a coarse graining for which the entropy of the rest is low in one time direction (the ``past").  These are  the subsystems allowing creatures that  ``live in time" ---such as those in the biosphere--- to exist.   I reply to some objections raised to an earlier presentation of this idea, in particular by Bob Wald, David Albert and Jim Hartle.

\end{abstract}

\maketitle
    
\section{Introduction}

An imposing aspects of the Cosmos is the mighty daily rotation of Sun, Moon, planets, stars and all galaxies around us.  Why does the Cosmos so rotate?  Well, it is not really the Cosmos to rotate, it is us. The rotation of the sky is a \emph{perspectival} phenomenon: we understand it better as due to the peculiarity of our own moving point of view, rather than a global feature of all celestial objects.  

A vivid feature of the world is its being in color:  each dot of each object has one of the colors out of a three-dimensional color-space.  Why?  Well, it is us that have \emph{three} kinds of receptors in our eyes, giving the 3d color space.  The 3d space of the world's colors is \emph{perspectival}: we understand it better as a consequence of the peculiarity of our own physiology, rather than the Maxwell equations. 

The list of conspicuous phenomena that have turned out to be perspectival is long; recognising them  has been a persistent aspect of the progress of science.

A vivid aspect of reality is the flow of time;  more precisely: the fact that the past is different from the future. Most observed phenomena violate time reversal invariance strongly. Could this be a perspectival phenomenon as well?   Here I suggest that this is a likely possibility.

Boltzmann's $H$-theorem and its modern versions show that for most microstates away from equilibrium, entropy increases in \emph{both} time directions \cite{Lanford1981,Uffink2010,Price,}.  Why then we observe lower entropy in the past? For this to be possible, most microstates around us appear to be very \emph{non} generic. This is the problem of the arrow of time, or the problem of the source of the second law of thermodynamics \cite{Lebowitz1993,Price}.  The common solution is to believe that the universe was born in an extremely non-generic microstate \cite{Carroll2010}. Roger Penrose even considered the possibility of a fundamental cosmological law breaking time-reversal invariance, forcing initial singularities to be extreemely special (vanishing Weil curvature) \cite{Penrose:1979fk}.  

Here I point out that there is a different possibility: past low entropy might be a \emph{perspectival} phenomenon, like the rotation of the sky. 

This is possible because entropy depends on the system's microstate \emph{but also} on the coarse graining under which the system is described. In turn, the relevant coarse graining is determined by the concrete existing interactions with the system. The entropy we assign to the systems around us depends on the way we interact with them ---as the apparent motion of the sky depends on our own motion.  

A subsystem of the universe that happens to couple to the rest of the universe via macroscopic variables determining an entropy that happens to be low in the past, is a system to which the universe  appears strongly time oriented.  As it appears to us.  Past entropy  may appear low because of our own perspective on the universe. 
 
Specifically, I argue below that the following conjecture is plausible: 
\begin{description} 
\item[Conjecture]  In a sufficiently complex system, there is always \emph{some} subsystem whose interaction with the rest determines a coarse graining with respect to which the system satisfies the second law of thermodynamics (in some time direction).  
\end{description}
An example where this is realized is given below. 

If this is correct, we have a new way for facing the puzzle of the arrow of time:  the universe is in a generic state, but is sufficiently rich to include subsystems whose coupling defines a coarse graining for wich entropy increases monotonically.  These subsystems are those where information can pile up and ``information gathering creatures" such as those composing the biosphere can exist.  

All phenomena related to time flow, all phenomena that distinguish the past from the future, can be traced to (or described in terms of) entropy increase. Therefore the difference between past and future  may follow from the peculiarities of our coupling to the rest of the universe, rather than from a peculiarity of the microstate of the universe. Like the rotation of the cosmos.

\section{A preliminary conjecture}

To start with, consider classical mechanics. Quantum theory is discussed in the last section.  It is convenient to use Gibbs' formulation of statistical mechanics rather than Boltzmann's, because Boltzmann  takes for granted the split of a system in a large number of equal subsystems (the individual molecules), and this may precisely offuscate the key point in the context of general relativity and quantum field theory, as we shall see. 

Consider a classical system with many degrees of freedom in a (``microscopic") state $s$, element of a phase space $\Gamma$, evolving in time as $s(t)$.  Let $\{A_n\}$, be a set of (``macroscopic") observables --real functions on $\Gamma$--,  labeled by the index $n$.   This set defines a coarse graining. That is, it partitions $\Gamma$ in unequal regions where the $A_n$ are constant. The largest of these regions is the equilibrium region. The entropy of a state $s$ can be defined as the volume of the region where it is located.  With a (suitably normalized and time invariant) measure $ds$, entropy is then 
\be
 S_{A_n}=\log \int_\Gamma ds' \prod_n \delta(A_n(s')-A_n(s)),
\ee
where the family of macroscopic observables $A_n$ is indicated in subscript to emphasise that the entropy depends on the choice of these observables.  Notice that this definition applies to any microstate.\footnote{This equation defines entropy up to an an additive factor, because phase space volume has the dimension of $[Action]^N$, where $N$ is the number of degrees of freedom. This  is settled by quantum theory, which introduces a unit of action, the Planck constant, whose physical meaning is to determine the minimum empirically distinguishable phase space volume, namely the maximal amount of information in a state. See \cite{Haggard:2013fx}.} 

 As the microstate $s$ evolves in time so does its entropy
\be
S_{A_n}(t)=\log \int_\Gamma ds' \prod_n \delta(A_n(s')-A_n(s(t))).
\ee  
Boltzmann's $H$-theorem and its modern versions imply that under suitable ergodic conditions  \emph{if we fix the choice of the macroscopic observables $A_n$}, for most microstates out of equilibrium at $t_0$, and for any finite $\Delta t$, we have  $S_{A_n}(t_0+\Delta t)>S_{A_n}(t_0)$ irrespectively of the sign of $\Delta t$. 

I want to bring the attention, instead, on the dependence of entropy on the family of observables, and enunciate the following first conjecture. If the system is sufficiently complex and ergodic, for most paths $s(t)$ that satisfy the dynamics and for each orientation of $t$, there is a family of observables $A_n$ such that 
\be
\frac{dS_{A_n}}{dt}\ge 0. 
\ee  
In other words, \emph{any} motion appears to have initial low entropy (and non decreasing entropy) under \emph{some} coarse graining. 

The conjecture become plausible with a concrete example. Consider a set $\Sigma$ of $N$ distinguishable balls that move in a box, governed by a time reversible ergodic dynamics. Let the box have an extension $x\in[-1,1]$ in the direction of the $x$ coordinate, and be ideally divided in two halves by $x\!=\!0$.  For any given subset  $\sigma\subset \Sigma$ of balls, define the observable $A_\sigma$ to be the mean value of the $x$ coordinate of the balls in $\sigma$.  That is, if $x_b$ is the $x$ coordinate of the ball $b$, define
\be
    A_\sigma = \frac{\sum_{b\in\sigma} x_b}{\sum_{b\in\sigma}1}.
\ee

Let $s(t)$ be a generic physical motion of the system, say going from $t\!=\!t_a$ to $t\!=\!t_b>t_a$.  Let $\sigma_a$ be the set of the balls that are at the right of $x\!=\!0$ at $t\!=\!t_a$.  The macroscopic observable $A_a\equiv A_{\sigma_a}$ defines an entropy that in the large $N$ limit and for most motions $s(t)$ satisfies
\be
\frac{S_{A_a}(t)}{dt}\ge 0. 
\ee
This is the second law of thermodynamics. 
\begin{figure}[t]
\includegraphics[width=4.2cm]{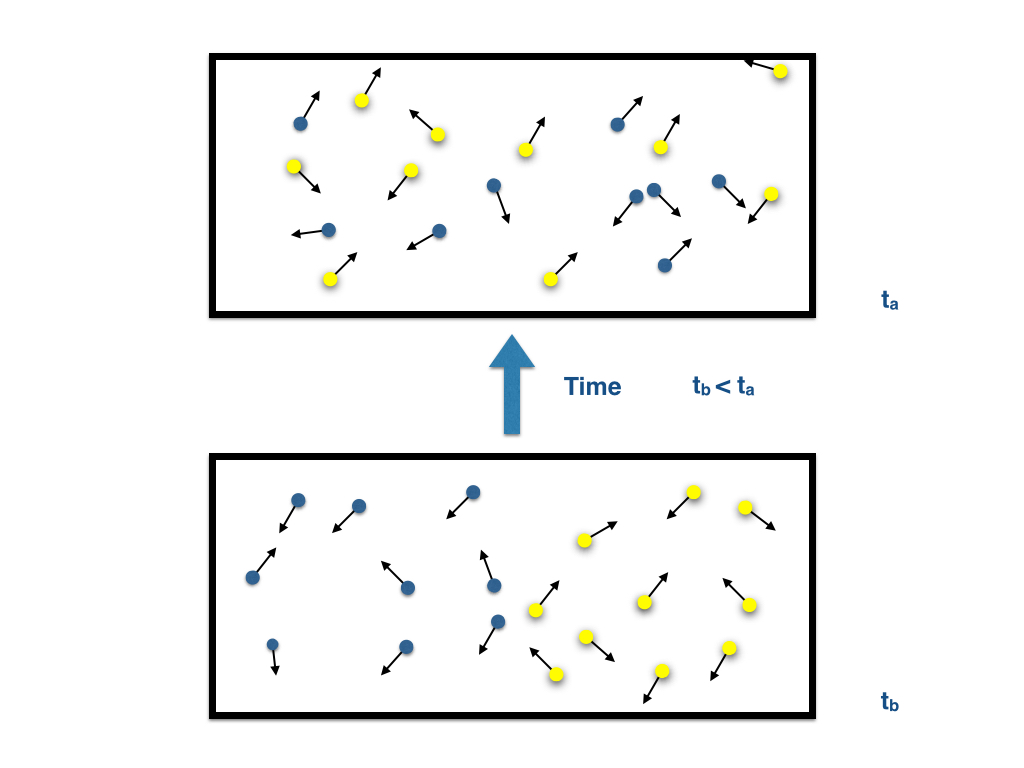}\ 
\includegraphics[width=4.2cm]{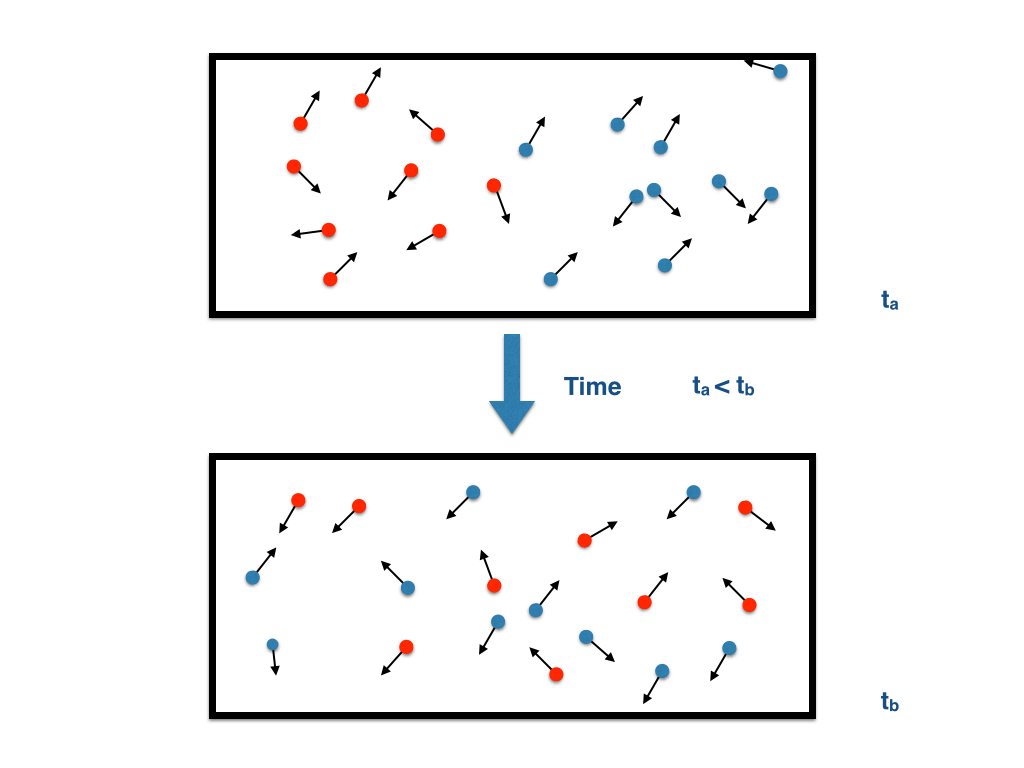}
\caption{The same history, seen with different filters: for a filter seeing the yellow balls that are on the right at time $t_a$, entropy is low at $t_a$. A filter that sees the red balls on the left at $t_b$ defines an entropy low at $t_b$. Since the direction of time flow is determined by increasing entropy, time flows in a different direction with respect to the two different observables.}
\end{figure}

But let's now fix the motion $s(t)$, and define a different observable as follows. Let $\sigma_b$ be the set of the balls that are at the left of $x\!=\!0$ at $t\!=\!t_b$.  The macroscopic observable $A_b \equiv  A_{\sigma_b}$ defines an entropy that is easily seen to satisfy
\be
\frac{S_{A_b}(t)}{dt}\le 0. 
\ee
This is again the second law of thermodynamics, but now in the reversed time $-t$.  It holds for the generic motion $s(t)$, with a specific observable.

This is pretty obvious: if at time $t_a$ we ideally color in yellow all the balls at the right of $x\!=\!0$ (See Figure~1), then the state at $t_a$ is low entropy with the respect to \emph{this} coarse graining, and the motion mixes the balls and raises the entropy as $t$ moves from $t_a$ to $t_b$.   But if instead we color in red the balls that are at the left of $x\!=\!0$ \emph{at the time $t_b$}, then the reverse is true and entropy increases in the reverse $t$ direction.  

The point is simple: for any motion there is a macroscopic family of observables with respect to which the state at a chosen end of the motion has low entropy: it suffices to choose observables that single out well the state at the chosen end.  I call these observables, ``time oriented". They are determined by the state itself.  

This simple example shows that, generically, past low entropy is not a feature of a special physical history of microstates of the system.  Each such histories may \emph{appear} to be time oriented (that is: have increasing entropy) under a suitable choice of macroscopic observables.

Can this observation be related to the fact that we see entropy increase in the world? An objection to this idea is: how can a physical fact of nature, such as the second law, depend on a choice of coarse graining, which ---so far--- seems subjective and arbitrary?  In the next section I argue that there is nothing arbitrary in the choice of the coarse graining and the macroscopic observables. These are fixed by the coupling between subsystems.   Different choices of coarse graining represent different possible subsystems. 

To pursue the analogy that opens this paper, different reference systems from which the universe can be observed are concretely realised by different rotating bodies, such as the Earth. 

\section{Time-oriented subsystems}

The fact that thermodynamics and statistical mechanics require a coarse graining, namely a ``choice" of macroscopic observables, appears at first sight to introduce a curious element of subjectivity into physics,  clashing with the objectivity of the predictions of science. 

But of course there is nothing subjective in thermodynamics. A cup of hot tea does not cool down because of what I know or do not know about its molecules. The ``choice" of macroscopic observables is dictated  by the ways the system under consideration couples.  The macroscopic observables of the system are those coupled to the exterior (in thermodynamics, those that can be manipulated and   measured). The thermodynamics and the statistical mechanics of a system defined by a set of macroscopic observables $A_n$ describes (objectively) the way the system interacts when coupled to another system via \emph{these} observables, and the way \emph{these} observables behave (a point emphasize recently for instance in \cite{Hemmo2012}. 

For instance, the behaviour of a box full of air is going to be described by a certain entropy function if the air is interacting with the exterior via a piston that changes its volume $V$.  But the \emph{same} air is going to be described by a \emph{different} entropy function if it interacts with the exterior via \emph{two} pistons with  filters  permeable to Oxygen and Nitrogen respectively. See Figure 2. In this case, the macroscopic observables are others and chemical potentials enters the game.   It is not our abstract ``knowledge" on the relative abundance of Oxygen and Nitrogen that matters: it is the presence or not of a  physical coupling of this quantity to the exterior, and the possibility of their independent variation, to determine which entropy describes the phenomena. Different statistics and thermodynamics of the same box of air describe different interactions of the box with the exterior. 

\begin{figure}
\includegraphics[width=7cm]{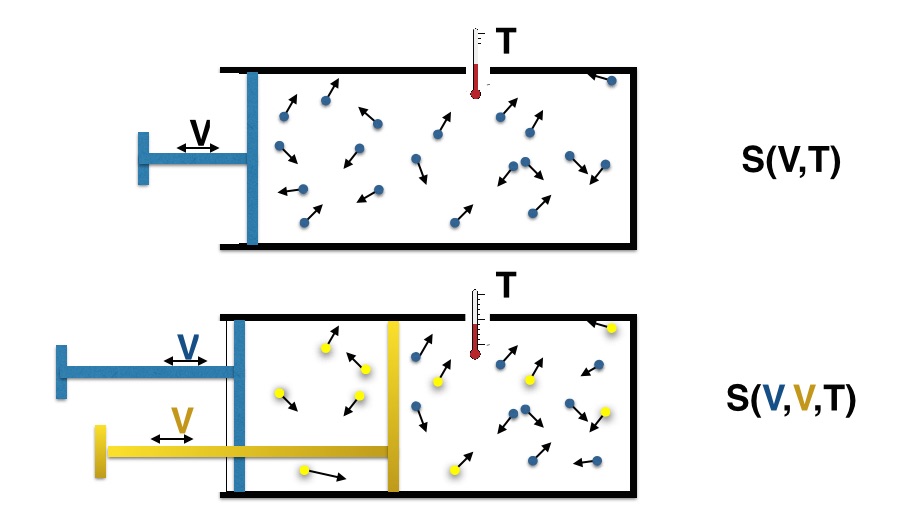}
\caption{The same system --here a volume of air-- is described by different entropy functions, describing different interactions it can have. Here via its total volume or via the volume of its distinct chemical  components.}
\end{figure}
In the light of this consideration, let us reconsider the box of the previous section replacing the abstract notion of ``observable" by a concrete interaction between subsystems. 

Say we have the $N$ balls in a box as above, but now we add a new set of $2^N$ ``small" balls\footnote{$2^N$ is the number of subsets of $\Sigma$, namely the cardinality of its power set.}, with negligible mass, that do not interact among themselves but interact with the previous (``large") balls as follows. Each small ball is labeled by a subset $\sigma\subset\Sigma$ and is attracted by the balls in $\sigma$ and only these, via a force law such that the total attraction is in the direction of the center of mass $A_\sigma$ of the balls in $\sigma$ (See Figure 3).

Generically, a small ball interacts with a large number of large balls, but it does so only via a single variable:  $A_\sigma$. Therefore it interacts with a statistical system, for which $A_\sigma$ is the single macroscopic observable. For each small ball $\sigma$, the ``rest of the universe"  behaves as a thermal system with entropy $S_{A_\sigma}$.  

It follows form the considerations of the previous sections that given a \emph{generic} motion $s(t)$ there will generically be at least one small ball,  the ball $\sigma_a$ for which the entropy of the rest of the box is never decreasing in $t$, in the thermodynamical limit of large $N$.  (There will also be another small ball, $\sigma_b$ for which the entropy of the rest of the box is never \emph{increasing} in $t$.)

Imagine that the box is the universe and each ``small" ball $\sigma$ is itself a large system with many degrees of freedom.  Then generically there is at least one of these, namely $\sigma_a$ (in fact many) for which the rest of the universe has a low-entropy initial state.  In other words, it is plausible to expect the validity of the conjecture stated in the introduction, which I repeat here:

\begin{description} 
\item[Conjecture]  In a sufficiently complex system, there is always \emph{some} subsystem whose interaction with the rest determines a coarse graining with respect to which the system satisfies the second law of thermodynamics (in some time direction).  
\end{description}

That is: low past entropy  can be fully perspectival. 

\begin{figure}
\includegraphics[width=8.5 cm]{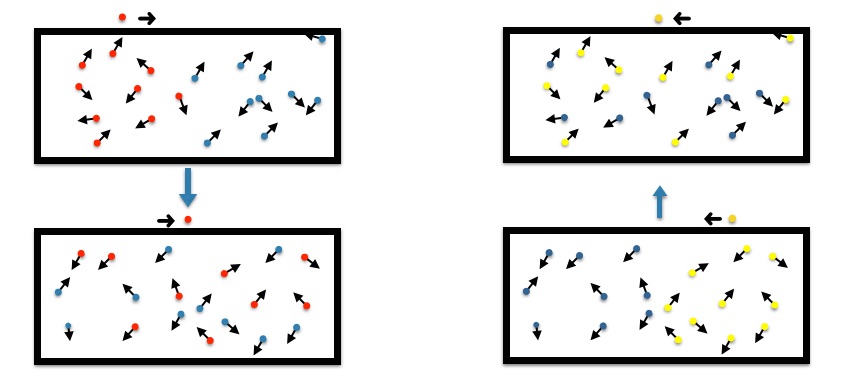}
\caption{The ``small" balls are represented on top of the box. The yellow and red ones are attracted, respectively, by the large yellow and red balls. Both interact with a statistical system where entropy changes, but entropy increases on the opposite direction with respect of each of them.}
\end{figure}

Now, since $\sigma_a$ interacts thermodynamically with a universe which was in a low-entropy state in the past, the world seen by $\sigma_a$ appears organized in time: observed phenomena display marked and consistent arrows of time, which single out one direction.  $\sigma_a$ interacts with a world where entropy increases, hence ``time flows" in one specific direction. The world seen by $\sigma_a$ may include dissipation, memory, traces of the past, and all the many phenomena that characterize the universe as we see it.  Within the subsystem $\sigma_a$, time oriented phenomena that require the growth of entropy, such as evolution or accumulation of knowledge, can take place.  I call such a subsystem a ``time oriented".

Could this picture be related to the reason why we see the universe as a system with a low-entropy initial state?  Could the low entropy of the universe characterize our own coupling with the universe, rather than a peculiarity of the microstate of the universe? 

In the next section I answer to a some possible objections to this idea. Some of these were raised and discussed at the 2015 Tenerife conference on the philosophy of cosmology. 

\section{Objections and replies}

{\em 1. Isn't this just shifting the problem? Instead of the mystery of a strange (low past entropy) microstate of the universe, we have now the new problem of explaining why we belong to a peculiar system?}   

Yes.  But it is easier to explain why the Earth happens to rotate, rather than  having to come up with a rational for the full cosmos to rotate.   The next question addresses the shifted problem. 

{\em 2. For most subsystems the couplings are such that entropy was not low at one end of time. Why then should \emph{we} belong to a special subsystem that couples in such a peculiar manner?}   

Because this is the condition for us to be what we are.  We live in time. Our own existence depends on the fact of being in a situation of strong local entropy production: biological evolution, biochemistry, life itself, memory, knowledge acquisition, culture...  As emphasized by David Albert, low past entropy is what allows us to reconstruct the past, and have memory and therefore gives us our sense of identity.  Being part of a time-oriented subsystem is the condition for all this.  This is a mild use of anthropic reasoning. It is analogous to asking why do we live on a planet's surface (non-generic place of the universe) and answering that this is simply what we are: creatures living on ground, needing water and so on.   Our inhabiting these quarters of the universe is no more strange than me being born in a place where people happen to speak my own language. 

{\em 3.  Assuming that we choose a coarse graining for which entropy is low at initial time $t_a$. Wouldn't then entropy move very fast  to a maximum, in the time scale of the molecular interactions, and then just fluctuate around the maximum? (Point raised by Bob Wald).} 

There are different time scales. The thermalisation time scale can be hugely different from the time scale of the molecular interactions, and in fact it is clearly so in our universe.   Given a history of a isolated system, a situation where entropy increases can exist only for a time scale shorter than the  thermalisation time.  This is precisely the situation in which we are in the universe: the Hubble time is much longer than the time scale of  microphysics, but much shorter than the thermalisation time of the visible universe.  So, the time scales are fine. 

{\em 4.  The interactions in the real universe are not as arbitrary as in the example of the heavy and small balls. In fact, in the universe there are no more than a small number of fundamental interactions. (Point raised by David Albert).} 

The fundamental interactions are only a few, but the interaction channels they open are innumerable.  The example of the colors makes this clear: the relevant elementary interaction is just the electromagnetic interaction.  But our eyes pick up an incredibly tiny component of the electromagnetic waves. They pick up three variables out of an infinite number: they recognise a three dimensional space of colors (with some resolution) out of the virtually infinite dimensional space of waveforms. So we are \emph{precisely} in the situation of the small balls, which only interact with a tiny fraction of the variables of the external world. It can be argued that these are the most relevant for us. This is precisely the point: it is by interacting with some specific variables that we may pick up time oriented features of the world.  Another simple example is a normal radio: it can easily tune on a single band, out of the entire electromagnetic spectrum.  We are in a similar situation. For instance, we observe a relatively tiny range of phenomena, among all those potentially existing  in the full range of time scales existing in the 60 orders of magnitude between the Planck time and cosmological time. 

{\em 5. We see entropy increase in cosmology, which is the description of the whole, without coarse graining.}

Current scientific cosmology is not the dynamics of everything: it is the description of an extremely coarse grained picture of the universe.  Cosmology is a feast of coarse graining.

{\em 6.  The observables that we use to describe the world are  coarse grained but they are the natural ones.} 

Too often ``natural" is just what we are used to. Considering something ``natural" is to be blind to subjectivity. For somebody it is natural to write from left to right. For others, the opposite.

{\em 7.  Our interactions pick up variables that are determined by the spatio-temporal structure of the world: spacetime integrals of conserved quantities. Quasi-classical domains are determined by these. Are these sufficiently generic for the mechanism you suggest? (Point raised by Jim Hartle).} 

Yes they are, because spacetime averages of conserved quantities  carry a very large amount of information, as our eyes testify. But this point is better raised in the context of quantum gravity, where the spacetime-regions structure is itself an emergent classical phenomenon that requires a quasi-classical domain. The emergence of a spatio-temporal structure from a quantum gravitational context may be related to the emergence of the second low in the sense I am describing here. In both cases there is a perspectival aspect of the emergence.

{\em 8.  Can the abstract picture of the coarse graining determining entropy be made concrete with an example?} 

The biosphere is an oriented subsystem of the universe. Consider the thermodynamical framework of life on Earth.  There is a constant flow of electromagnetic energy on Earth: incoming radiation from the Sun and outgoing radiation towards the sky. Microscopically, this is a certain complicate solution of Maxwell equation.  But as far as life is considered, most details of this solution  (such as the precise phase of a single solar photon falling on the Pacific ocean) are irrelevant.

What matters to life on Earth are energy and a certain range of frequency, integrated over small regions. This determines a coarse graining and therefore a notion of entropy. Now the incoming energy is the same as the outgoing energy, but not so for the frequency.  The Earth receives energy  $E$ (from the Sun) at higher frequency $\nu_a$ and emits energy (towards the sky) at lower frequency $\nu_b$ (See Figure 4).   This is a fact about the actual solution of the Maxwell equations in which we happen to live.  If we take energy and frequency as macroscopical observables, then an entropy is defined by such coarse graining.  Roughly, this entropy counts the number of photons; at frequency $\nu$ the number of photons in a wave of energy $E$ is $N\!=\!E/\hbar\nu$. If the received energy is emitted at lower frequency, the emitted entropy $S_b$ is higher than the received entropy $S_a$.  The process produces entropy: $S_b\gg S_a$.  This entropy production is not a feature of the  solution of the Maxwell equations alone:  it is a feature of this solution \emph{and} a set of macroscopic observables (integrated energy and frequency: oriented observables for this solution of the Maxwell equation) to which living systems couple.  

\begin{figure}
\includegraphics[width=6cm]{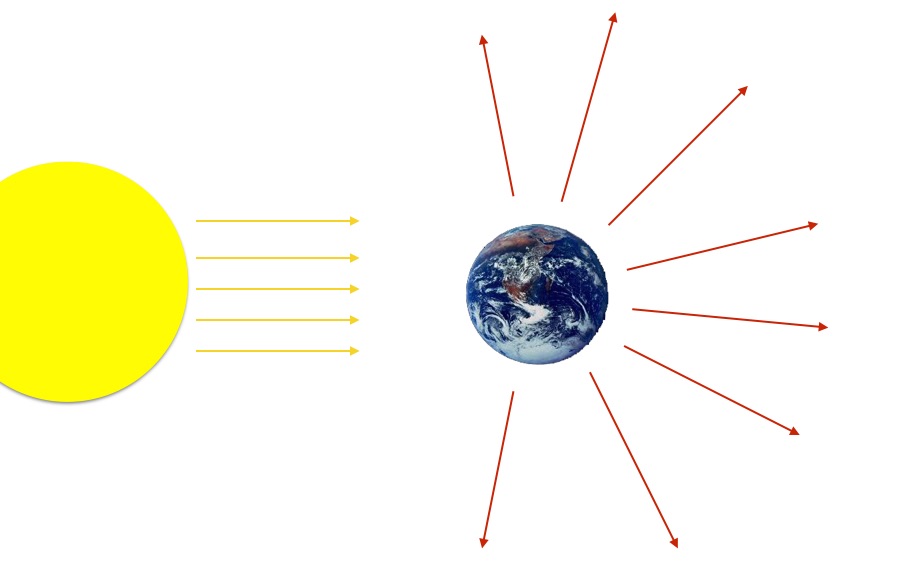}
\caption{Electromagnetic energy enters and exit the Earth. The biosphere interact with coarse grained aspects of it (frequency) with reset to which there is entropy production, and therefore time orientation, on the planet.}
\end{figure}

Any system on Earth whose dynamics is governed by interactions with $E$ and $\nu$ has a source of negative entropy at its disposal.  This is what is exploited by the biosphere on Earth to build structure and organization. The point I am emphasising is that what is relevant and peculiar here is not the individual solution of the Maxwell equation describing the incoming and outgoing waves: it is the peculiar manner in which the interaction with this energy is coarse grained by the biosphere.

\section{Quantum theory and general relativity}

Quantum phenomena provide a source of entropy distinct from the classical one generated by coarse graining: entanglement entropy. The state space of any quantum system is described by a Hilbert space $\H$, with a  linear structure that plays a major role for physics.  If the system can be split into two components, its state space splits into the tensor product of two Hilbert spaces: $\H\!=\!\H_1\otimes\H_2$, each carrying a  subset of observables. Because of the linearity, a generic state is not a tensor product of component states; that is, in general $\psi\ne\psi_1\otimes\psi_2$. This is entanglement. Restricting the observables to those of a subsystem, say system $1$, determines a quantum entropy over and above classical statistical entropy. This is measured by the von Neumann entropy $S\!=\!-\tr[\rho\log\rho]$ of the density matrix $\rho\!=\!\tr_{\H_2}|\psi\rangle\langle\psi|$.  Coarse graining is given by the restriction to the observables of a single subsystem. 

The conjecture presented in this paper can then be extended to the quantum context. Consider a ``sufficiently complex" quantum system.\footnote{This means: with a sufficient complex algebra of observables and a Hamiltonian which is suitably ``ergodic" with respect to it. A quantum system is not determined uniquely by it Hilbert space, Hamiltonian and state. All separable Hilbert space are isomorphic, and the spectrum of the Hamiltonian, which is the only remaining invariant quantity, is not sufficient to characterise the system.} Then:
\begin{description} 
\item[Conjecture]  Given a generic state evolving in time as $\psi(t)$, there exists splits of the system into subsystems such that the von Neumann entropy is low at initial time and increases in time.\footnote{and others for which entropy is low at final time.}  
\end{description}
The point here is to avoid assuming a fixed tensorial structure of $\H$ a priori. Instead, given a generic state, we can find a tensorial split of $\H$ which sees von Neumann entropy grow in time. 

This conjecture, in fact, is not hard to prove.  A separable Hilbert space admits many discrete bases $|n\rangle$. Given any $\psi\in\H$, we can always choose a basis $|n\rangle$ where $\psi\!=\!|1\rangle$. Then we can consider two Hilbert spaces,  $\H_1$ and $\H_2$, with bases $|k\rangle$ and $|m\rangle$, and map their tensor product to $\H$ by identifying  $|k\rangle \otimes |m\rangle$ with the state $|n\rangle$ where $(k,m)$ appear, say, in the $n$-th position of the Cantor ordering of the $(n,m)$ couples  ((1,1),(1,2),(2,1),(1,3),(2,2),(3,1),(1,4)...).  Then, $\psi\!=\! |1\rangle \otimes |1\rangle$ is a tensor state and has vanishing von Neumann entropy. On the other hand, recent results show that entanglement entropy generically evolve towards maximizing entropy of a fixed tensor split (see \cite{Deutsch2013a}). 

Therefore for any time evolution $\psi(t)$ there is a split of the system into subsystems such that the initial state has zero entropy and then entropy grows.  Growing and decreasing of (entanglement) entropy is an issue about how we split the universe into subsystems, not a feature of the overall state of things (on this, see  \cite{Tegmark2012a}). Notice that in quantum field theory there is no single natural tensor decomposition of the Fock space.
 
Finally, let me get to general relativity. In all examples above, I have considered non-relativistic systems where a notion of the single time variable is clearly defined. I have therefore discussed the \emph{direction} of time, but not the \emph{choice} of the time variable.  In special relativity, there is a different time variable for each Lorentz frame.  In general relativity, the notion of time further breaks into related but distinct notions, such as proper time along worldliness, coordinate time, clock time, asymptotic time, cosmological time...  Entropy increase becomes a far more subtle notion, especially if we take into account the possibility that thermal energy leaks to the degrees of freedom of the gravitational field and therefore macrostates can includes microstates with different spacetime geometries. In this context, a formulation of the second law of thermodynamics requires to identify not only a  \emph{direction} for the time variable, but also the \emph{choice} of the time variable itself in terms of which the law can hold \cite{Josset2015}.  In this context, a spit of the whole system into subsystems is even more essential than in the non-relativistic case, in order to understand thermodynamics  \cite{Josset2015}. The observation mad in this paper therefore apply naturally to the non relativistic case. 

\vspace{.2cm}

The perspectival origin of many aspects of our physical world has been recently emphasised by some of the philosophers most sensible to modern physics \cite{Ismael2007,Price2011}.  I believe that the arrow time is not going to escape the same fate. 

The reason for the entropic peculiarity of the past Êshould not be sought in the cosmos at large. ÊThe place to look for them is in the split, and therefore in the macroscopic observables that are relevant to us. Time asymmetry, and therefore ``time flow", might be a feature of a subsystem to which we belong,  features needed for information gathering creatures like us to exist,  not a feature of the universe at large. 

\centerline{------}  

I am indebted to Hal Haggard and Angelo Vulpiani for useful exchanges and to Kip Thorne, David Albert, Jim Hartle, Bob Wald and several other participants to the 2014 Tenerife conference on the Philosophy of Cosmology for numerous inputs and conversations. 

\providecommand{\href}[2]{#2}\begingroup\raggedright\endgroup

\end{document}